\documentstyle[twoside, epsf]{article}

\input ibvs2.sty

\begin{document}
\IBVShead{6188}{10 November 2016}

\IBVStitle{An investigation of the RCB star candidate GDS\,J0702414-023501}

\IBVSauth{H{\"U}MMERICH, S.$^{1,2}$; BERNHARD, K.$^{1,2}$}

\IBVSinst{Bundesdeutsche Arbeitsgemeinschaft f{\"u}r Ver{\"a}nderliche Sterne e.V. (BAV), Berlin, Germany; \\ e-mail: ernham@rz-online.de}
\IBVSinst{American Association of Variable Star Observers (AAVSO), Cambridge, USA}

\SIMBADobj{IRAS 07001-0230}
\IBVStyp{ RCB, M }
\IBVSkey{stars: variables: RCB; variables: Mira}
\IBVSabs{2MASS J07024146-0235017 = GDS\,J0702414-023501 was included in the "Catalogue enriched with R CrB }
\IBVSabs{stars" on grounds of its near- and mid-infrared colours. The object, which corresponds to the} 
\IBVSabs{carbon star IRAS 07001-0230 = CGCS 6197, has been found to exhibit large amplitude variability}
\IBVSabs{in its Bochum Galactic Disk Survey light curve. Taking into account all available data,} 
\IBVSabs{GDS_J0702414-023501 is here proposed as a new candidate Mira variable.}

\begintext

R Coronae Borealis (hereafter RCB) stars are hydrogen-deficient, carbon-rich supergiant stars. Their photometric variability is characterised by unpredictable fading events (up to $\sim$8 magnitudes ($V$)), which are thought to be caused by the formation of carbon dust (e.g. Clayton 2012). RCB stars have been shown to possess warm and bright circumstellar shells, which are readily detected at mid-infrared wavelengths. Therefore, near- and mid-infrared colour-colour diagrams and cuts can be employed as a viable and efficient method of identifying new RCB candidates (cf. e.g. Feast (1997); Alcock et al. (2001); and Tisserand (2012)).

Using the above mentioned colour selection criteria, the star 2MASS J07024146-0235017 was included into the 'Catalogue enriched with R CrB stars' by Tisserand (2012). No time series photometry had been available for this object until the advent of the Bochum Galactic Disk Survey (GDS hereafter), which has been monitoring stars in a 6\deg wide strip along the Galactic plane and comprises photometry for stars in the magnitude range $8<r'<18$\,mag and $7<i'<17$\,mag. For more information on the GDS, the reader is referred to Haas et al. (2012) and Hackstein et al. (2015).

Hackstein et al. (2015) present a sample of 64\,151 variable sources identified in the GDS. Among them is 2MASS J07024146-0235017, which is listed under the designation GDS\,J0702414-023501 with a mean magnitude of 14.15 mag ($i'$) and an amplitude of 4.96 mag ($i'$). No variability type was proposed by the aforementioned authors. Basic data and archival photometry of GDS\,J0702414-023501 are presented in Table 1; positional information was derived from the 2MASS catalogue (Skrutskie et al. 2006).

\begin{table}
\caption{Table 1. Basic data and archival photometry of GDS\,J0702414-023501.}
\begin{center}
\begin{tabular}{ll}
\hline
ID & GDS\,J0702414-023501, 2MASS J07024146-0235017,\\
 & IRAS 07001-0230, CGCS 6197 \\
pos (RA, Dec; J2000) & 07\hr02\mm41\fsec461, $-$02\deg35\arcm01\farcs76 \\
$J$, $H$, $K_{s}$ (2MASS) & 11.957\,mag, 9.417\,mag, 7.402\,mag \\
$W1$, $W2$, $W3$, $W4$ (WISE) & 5.075\,mag, 3.124\,mag, 2.193\,mag, 1.563\,mag \\ \relax
[9$\mu$m], [18$\mu$m] (AKARI) & 8.509\,Jy, 3.101\,Jy \\ \relax
[12$\mu$m], [25$\mu$m], [60$\mu$m] (IRAS) & 5.045\,Jy, 2.387\,Jy, 0.548\,Jy \\
\hline
\end{tabular}
\end{center}
\end{table}

We have investigated the object because of its inclusion in the Tisserand (2012) catalogue and the observed, large-amplitude variability in GDS data. The light curve based on the current data release (Hackstein et al. 2015) is shown in Figure 1; only $i'$ data are available for this object. Although there are only 71 data points scattered over a time span of $\sim$1600 days, it becomes obvious that the star's range of variability is large (approximate range of 12.5\,mag\,$<$\,$i'$\,$<$\,17\,mag). Furthermore, the light curve is reminiscent of an RCB variable and seems to provide evidence against a Mira or RV Tauri classification -- two types of variables that, according to our experience, contaminate the RCB-enriched catalogue of Tisserand (2012). In particular, the light curve indicates that the star's brightness has been more or less constant for about 160 days around HJD 2456300, hovering around a mean magnitude of $\sim$16.5 mag ($i'$). Because of the available data, GDS\,J0702414-023501 has been listed as an RCB candidate (variability type RCB:) in the AAVSO International Variable Star Index (VSX; Watson 2006).

\IBVSfig{8cm}{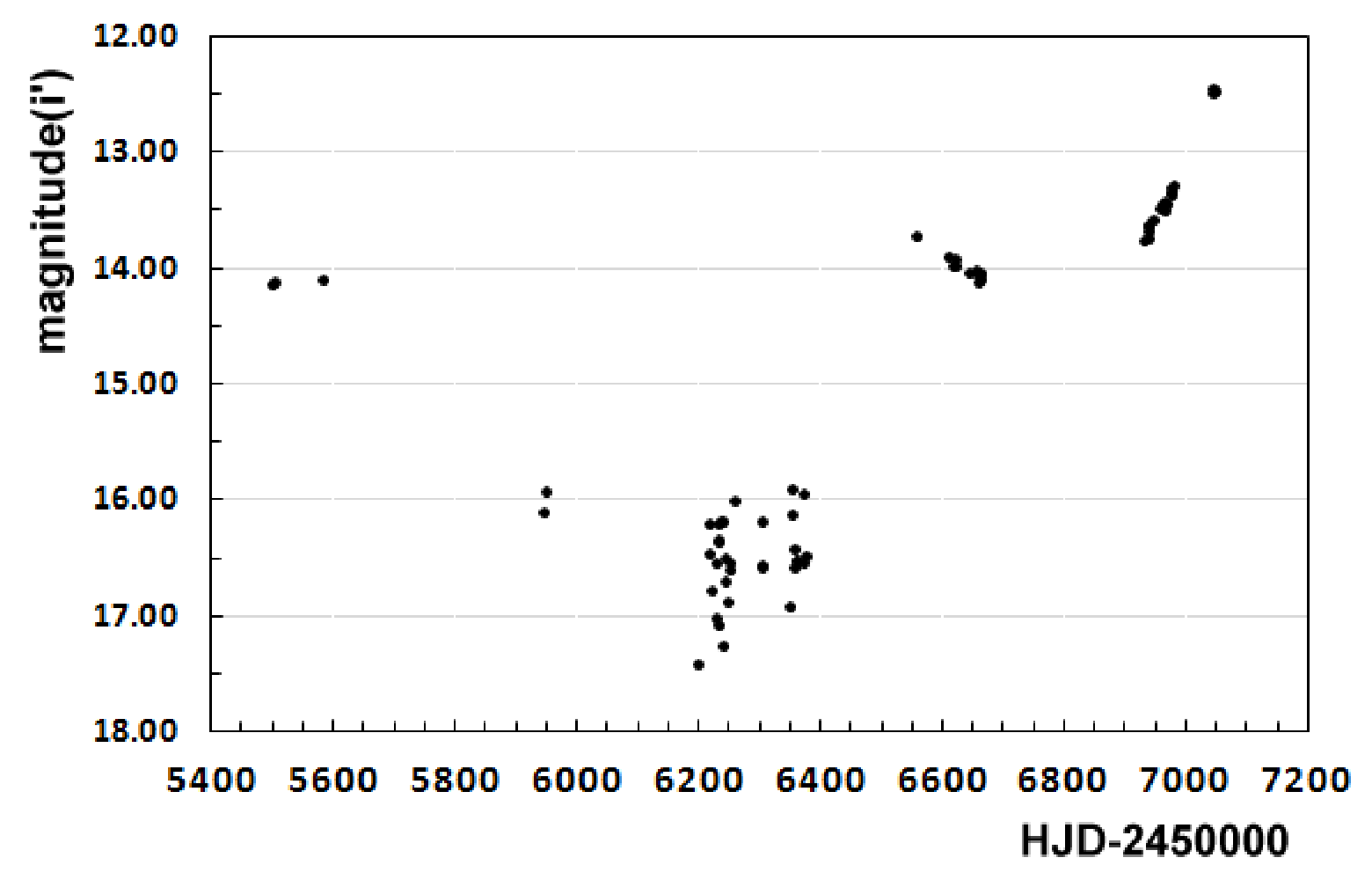}{The GDS $i'$ light curve of GDS\,J0702414-023501, based on the current data release (Hackstein et al. 2015).}
\IBVSfigKey{6188-f1.ps}{GDS J0702414-023501}{light curve}

We have searched sky survey plate archives for the existence of exposures of the sky region of our interest (Figure 2). Unfortunately, no images are available that show the star in a very bright state, as recorded by the GDS. Nevertheless, the star is considerably brighter on $I$-band sky survey plates from 1997 than it was in 1985 (two rightmost panels in Figure 2).

Based on the calculations of Schlafly and Finkbeiner (2011), we estimate an interstellar extinction of $A_{V}\approx2.2$ mag and $E_{B-V}\approx0.7$ mag for the sky area of our interest. Dereddened magnitudes in the 2MASS $J$, $H$, $K{\rm s}$ and WISE $W1$ and $W2$ bands (Cutri et al. 2012) were derived from the aforementioned source, which were used in the construction of the following colour-colour plots.

\IBVSfig{4cm}{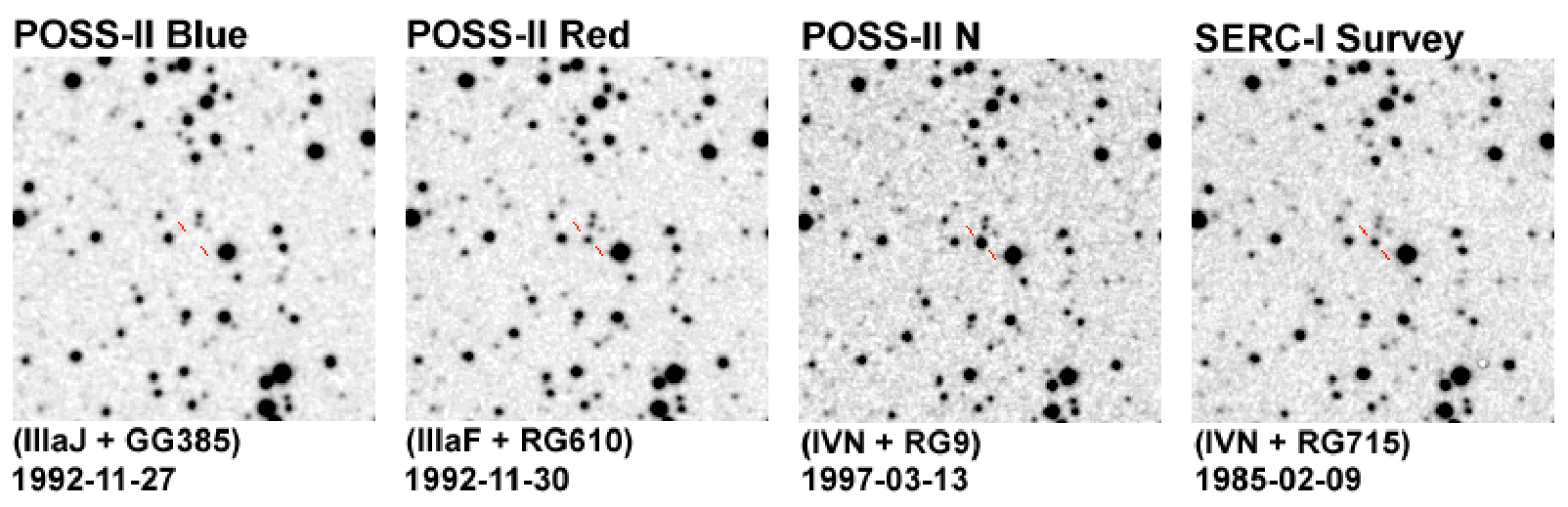}{GDS\,J0702414-023501 on archival sky survey plates (position of the star is indicated). The captions provide information on emulsions and filters used, as well as the epoch the plate was taken.}
\IBVSfigKey{6188-f2.ps}{GDS J0702414-023501}{finding chart}

A $(J-H)$ vs. $(H-K_{\rm s})$ diagram is shown in Figure 3. The RCB candidate GDS\,J0702414-023501 is indicated by the red and green dots which denote the loci of the star based on, respectively, reddened and unreddened magnitudes. Also shown are several confirmed RCB variables (open symbols), Mira variables (filled symbols) and the RV Tauri star UY Ara (cross). Classifications were taken from the GCVS (Samus et al. 2007-2016) and VSX online databases; data were drawn from the 2MASS catalogue.

The Mira variables approximately follow the positions of SMC carbon stars (solid line), as computed by Westerlund et al. (1991) and employed in this particular context by Tisserand et al. (2004). The very red, carbon-rich Mira V831 Mon is roughly situated on an extension of this line. The RCB stars, however, follow the dashed line, which illustrates the loci of a combination of two blackbodies representing the photosphere of the star ($\sim$5500 K) and the dust shell ($\sim$900 K), as employed by Feast (1997). The represented flux ranges from `all star' (lower, left end) to `all shell' (upper, right end).

\IBVSfig{6cm}{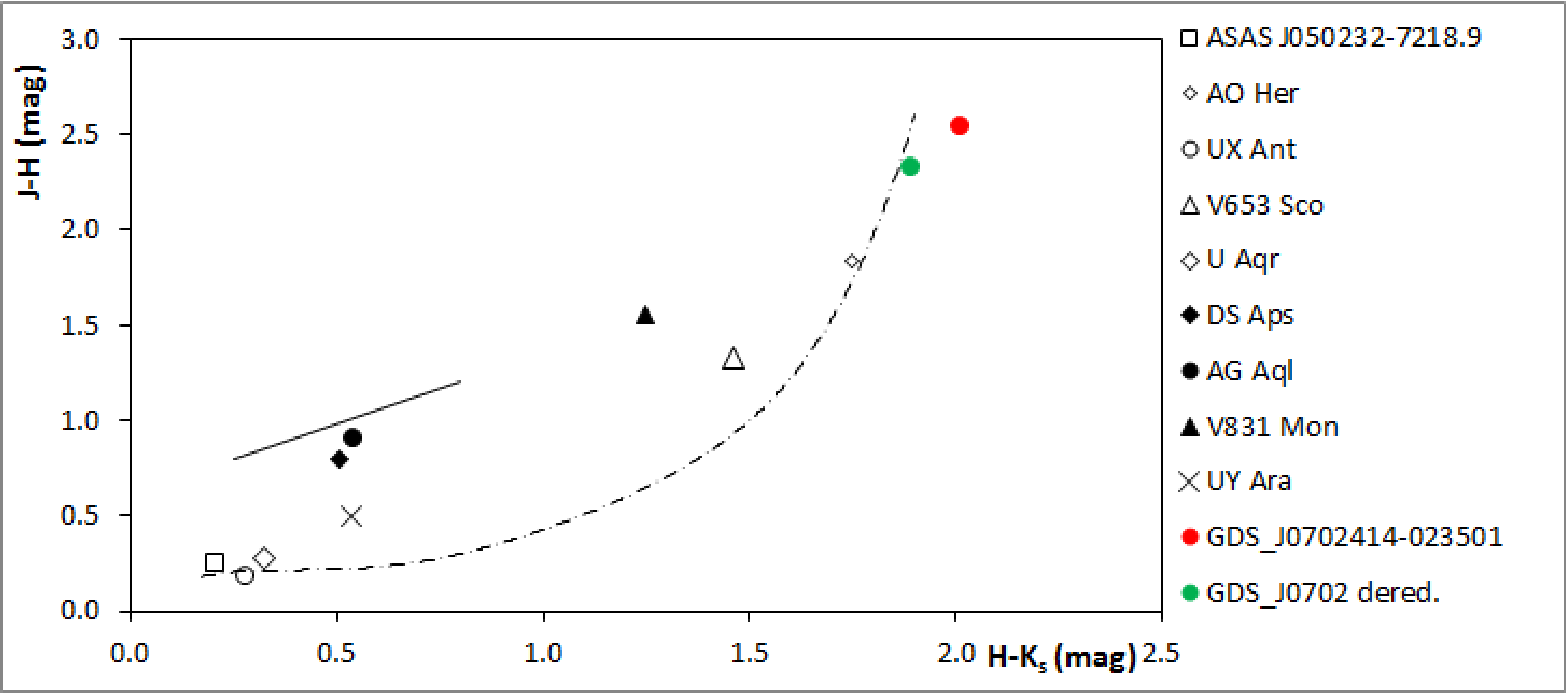}{$(J-H)$ vs. $(H-K_{\rm s})$ diagram, indicating the positions of GDS\,J0702414-023501 (red and green dots) and several confirmed RCB variables (open symbols), Mira variables (filled symbols) and the RV Tauri star UY Ara (cross). See text for details.}
\IBVSfigKey{6188-f3.ps}{GDS J0702414-023501}{other}

The position of GDS\,J0702414-023501 is reminiscent of the positions of RCB variables during deep obscuration minima (cf. e.g. the position of UW Cen during deep minima, indicated by the small crosses in Fig. 1 of Feast (1997)).

\IBVSfig{6cm}{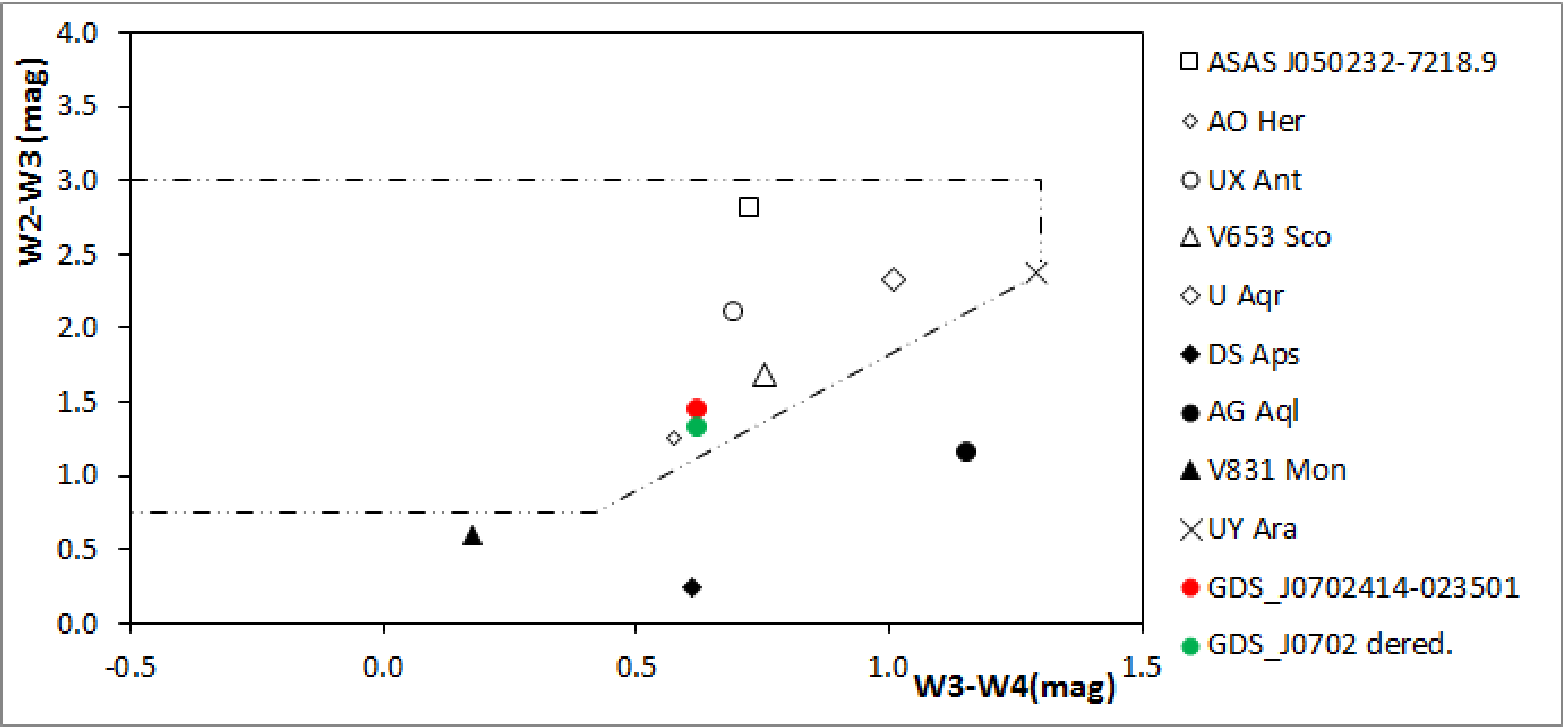}{$(W2-W3)$\,vs.\,$(W3-W4)$ diagram, indicating the positions of GDS\,J0702414-023501 (red and green dots) and several confirmed RCB variables (open symbols), Mira variables (filled symbols) and the RV Tauri star UY Ara (cross). See text for details.}
\IBVSfigKey{6188-f4.ps}{GDS J0702414-023501}{other}

Figure 4 shows the $(W2-W3)$ vs. $(W3-W4)$ diagram and has been based on WISE data. Stars and symbols are the same as in Figure 3. Selection cut (1) of Tisserand (2012), which effectively identifies objects with a shell signal, is indicated by the dashed lines. Only the confirmed RCB variables and GDS\,J0702414-023501 are well inside the indicated area.

\IBVSfig{7cm}{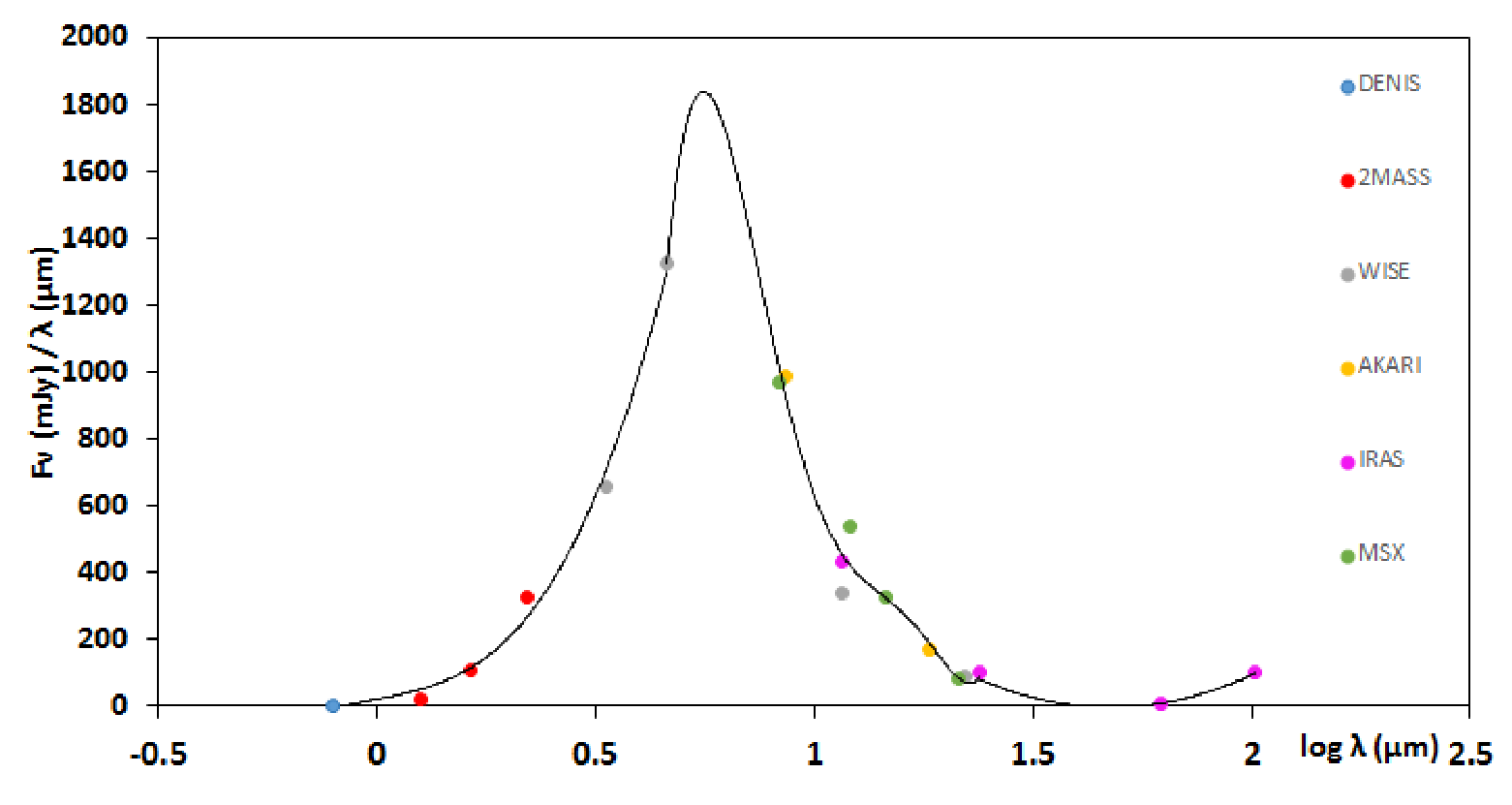}{Spectral energy distribution of GDS\,J0702414-023501, based on data from various catalogues, as indicated in the inset legend. The solid line indicates a polynomial fit of 6th order.}
\IBVSfigKey{6188-f5.ps}{GDS J0702414-023501}{other}

The spectral energy distribution (SED) is shown in Figure 5, which has been based on data obtained with the VizieR Photometry viewer.\footnote{\tt  http://vizier.u-strasbg.fr/vizier/sed/} The data have not been corrected for line-of-sight extinction. Note that the IRAS value at 100 $\mu$m denotes an upper limit. An IR excess due to the presence of warm dust is clearly visible. However, the SED differs from that of typical RCB stars in that there is no indication of cold dust (cf. e.g. Fig. 6 of Tisserand et al. 2009). This does not necessarily exclude an RCB classification, as the amount and temperature of dust around the star is related to the frequency and duration of obscuration events. However, it casts doubt on the proposed variability type.

The IRAS source 07001-0230 lies 15\farcs1 distant from the 2MASS position of \linebreak GDS\,J0702414-023501. Interestingly, IRAS 07001-0230 was identified as a carbon star on ground of its infrared colours by Guglielmo et al. (1993) and later entered the General Catalogue of Carbon Stars as CGCS 6197 (Alksnis et al. 2001). The identification of IRAS 07001-0230 with 2MASS J07024146-0235017 = GDS\,J0702414-023501 seems secure (MacConnell 1993; Chen et al. 2012). In MacConnell (1993), a $V$ magnitude of ~20.0 mag is indicated for IRAS 07001-0230, which elicits the following remark from B. Skiff: ``CGCS 6197, too faint (mi=16.7)". This may be taken as a hint at the large amplitude of variability observed in GDS\,J0702414-023501, which has been found as bright a 12.5 mag ($i'$).

\IBVSfig{10cm}{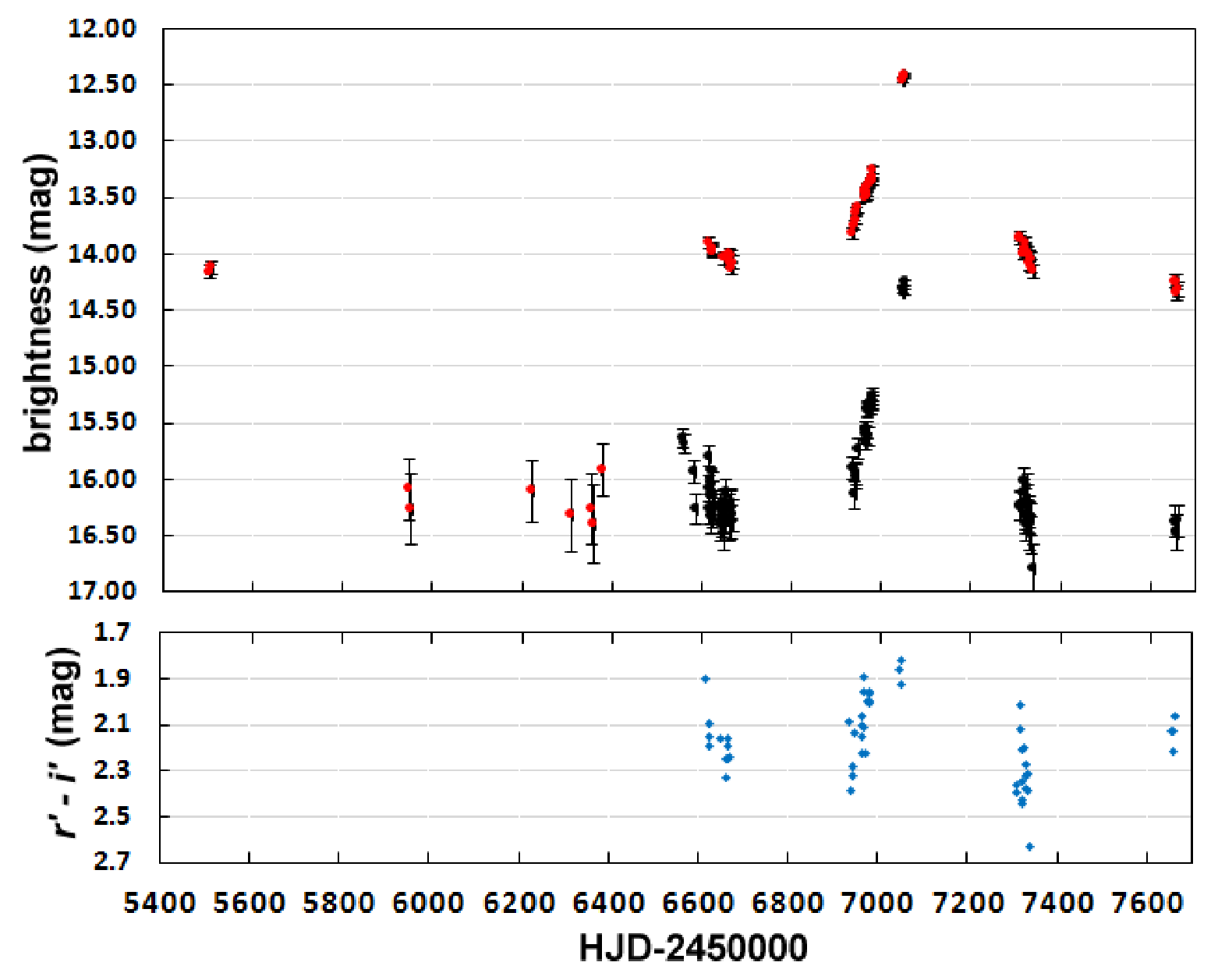}{The GDS light and colour curves of GDS\,J0702414-023501, based on recent, newly reduced and as yet unpublished GDS data (M. Hackstein, private communication). Red and black dots indicate, respectively, GDS $i'$ and $r'$ data.}
\IBVSfigKey{6188-f6.ps}{GDS J0702414-023501}{light curve}

In order to shed more light on the nature of our object of interest, new and hitherto unpublished GDS data were procured (M. Hackstein, private communication), which have been based on an improved reduction procedure and extend the time baseline by $\sim$600 days (Figure 6). Furthermore, the new data also boast measurements in $r'$, thereby allowing an investigation of the colour index evolution (Figure 6, bottom panel).

The new data indicate that about 300 days after the rise to $i' \sim$ 12.5 mag (at which point the hitherto available dataset terminates), a rather sharp drop in brightness takes place. Another $\sim$300 days later, several datapoints suggest a further drop or rise in brightness. Moreover, the new and improved data reduction procedure significantly reduced the number of faint datapoints around HJD 2456300; what formerly looked like a phase of approximately constant brightness (cf. Figure 1) is now down to some scattered data points and the light curve shape during this part is open to conjecture.

Generally, RCB stars show very diverse and sometimes peculiar light curves. However, the new GDS light and colour curves are strongly reminiscent of a Mira-type variable. Although the star has passed the selection criteria of Tisserand (2012) and is a confirmed carbon star, an RCB classification seems therefore unlikely; as has been pointed out above, Mira variables are known to contaminate the candidate list of Tisserand (2012).

Taking into account all available data, we propose GDS\,J0702414-023501 as a long-period variable of the Mira type. However, further photometric and spectroscopic studies are needed to reach a final conclusion.

\vskip5mm

{\noindent\bf Acknowledgements:} We thank Moritz Hackstein (Ruhr-Universit{\"a}t Bochum, Germany) for providing us with the most recent and newly reduced GDS observations, and the referee for helpful suggestions and valuable advice. This research has made use of the VizieR database, operated at CDS, Strasbourg, France.

\references

Alcock, C., et al., 2001, {\it ApJ}, {\bf 554}, 298 

Alksnis, A., et al., 2001, {\it BaltA}, {\bf 10}, 1 

Chen, P.S., et al., 2012, {\it AJ}, {\bf 143}, 36 

Clayton, G.C., 2012, {\it JAVSO}, {\bf 40}, 539 

Cutri, R.M., et al., 2012, {\it WISE All-Sky Data Release}, VizieR Online Data Catalogue ({\tt http://cdsarc.u-strasbg.fr/viz-bin/Cat?II/311})

Feast, M.W., 1997, {\it MNRAS}, {\bf 285}, 339 

Guglielmo, F., 1993, {\it A\&AS}, {\bf 99}, 31 

Haas, M., et al., 2012, {\it AN}, {\bf 333}, 706 

Hackstein, M., et al., 2015, {\it AN}, {\bf 336}, 590 

MacConnell, D.J., 1993, VizieR Online Data Catalogue \\({\tt http://cdsarc.u-strasbg.fr/viz-bin/VizieR?-source=III/170B})

Samus, N.N., et al., 2007-2016, {\it General Catalogue of Variable Stars}, VizieR Online Data Catalogue ({\tt http://cdsarc.u-strasbg.fr/viz-bin/Cat?B/gcvs})

Schlafly, E.F., \& Finkbeiner, D.P., 2011, {\it ApJ}, {\bf 737}, 103 

Skrutskie, M.F., et al., 2006, {\it AJ}, {\bf 131}, 1163 

Tisserand, P., et al., 2004, {\it A\&A}, {\bf 424}, 245 

Tisserand, P., et al., 2009, {\it A\&A}, {\bf 501}, 985 

Tisserand, P., 2012, {\it A\&A}, {\bf 539}, A51 

Watson, C.L., 2006, {\it SASS}, {\bf 25}, 47 

Westerlund, B.E., et al., 1991, {\it A\&AS}, {\bf 91}, 425 

\endreferences

\end{document}